# Effect of Fast Scale Factor Fluctuations on Cosmological Evolution

**Igor I. Smolyaninov**

Department of Electrical and Computer Engineering, University of Maryland, College Park, MD 20742, USA; smoly@umd.edu

**Abstract:** In this paper we study the corrections to the Friedmann equations due to fast fluctuations of the universe scale factor. Such fast quantum fluctuations were recently proposed as a potential solution of the cosmological constant problem. They also induce strong changes to the current sign and magnitude of the average cosmological force, thus making it one of the potential probable causes for the modification of Newtonian dynamics in galaxy-scale systems. It appears that quantum fluctuations of the scale factor also modify the Friedmann equations, leading to considerable modification of cosmological evolution. In particular, they give rise to late-time accelerated expansion of the universe, and they may also considerably modify the effective universe potential.

## 1. Introduction

It is widely believed that quantum fluctuations of matter and fields play a significant role in the cosmological evolution by defining the temporal behavior of the universe scale factor $a(t)$ resulting from the Friedmann equations. Very recently Wang et al. suggested that very fast quantum fluctuations of the universe scale factor may also considerably alter the universe dynamics on very fast time scales, thus providing a potential solution to the cosmological constant problem [1]. The approach developed in [1] implies that at the ultra-fast and microscopic (Planck) scales, the "true" magnitude of the cosmological constant is very large, and its magnitude is defined by the vacuum energy density predicted by quantum mechanics, which is about 120 orders of magnitude larger than the value observed in the experiment at cosmological scales. However, the effect of this huge "true" cosmological constant at macroscopic scales is largely canceled by strong metric fluctuations on the Planck scale. As a result, the effective cosmological constant observed at macroscopic cosmological scales equals its currently measured value of $\Lambda=1.11 \times 10^{-52}$ m$^{-2}$ [2]. Cosmological solutions exhibiting fast scale factor oscillations also appear quite generically in many other theoretical frameworks, such as semiclassical quantum gravity [3], cosmological models with oscillating dark energy [4], and various modified gravity theories [5]. It should be noted that the role of the universe scale factor $a(t)$ is somewhat similar to the role of the center of mass coordinate, which describes motion of some complicated macroscopic object. In classical mechanics this motion is well defined. However, in the quantum mechanical picture the center of mass motion obeys the uncertainty principle, and it is not well-defined anymore. Like any other object, the center of mass of a system experiences quantum fluctuations. Similar to quantum mechanics, in the semi-classical quantum gravity picture, we should expect $a(t)$ not to be well-defined. Instead, the universe scale factor $a(t)$ must experience quantum fluctuations, especially since there is no reason to expect that the quantum state of the universe within the scope of the quantum Friedmann model will always be a scale factor eigenstate.

Experimental evidence of relatively fast scale factor oscillations also started to emerge very recently [6]. In addition, it was demonstrated that fast scale factor oscillations modify the current sign and magnitude of the average cosmological force, thus making it one of the potential probable causes for the modification of Newtonian dynamics in galaxy-scale systems [7]. Therefore, it may be also interesting to consider the effect such fast fluctuations of the cosmological scale factor might play in the macroscopic evolution of the universe on the slow cosmological scales. We will demonstrate that quantum fluctuations of the scale factor modify the Friedmann equations, giving rise to late-time accelerated expansion of the universe, and they may also considerably modify the effective universe potential.

## 2. Materials and Methods

Let us describe the spatially homogeneous and isotropic universe using the standard Friedmann–Lemaître–Robertson–Walker (FLRW) metric:

$$ds^2 = -c^2 dt^2 + a(t)^2 ds_3^2, \tag{1}$$

where $a(t)$ is the cosmological scale factor, and $ds_3^2$ is a three-dimensional metric having a constant (positive, zero, or negative) curvature. The two Friedmann equations, which define the cosmological evolution of $a(t)$ may be written as

$$\frac{\dot{a}^2}{a^2} = \frac{8\pi}{3}G\rho + \frac{\Lambda c^2}{3} - \frac{kc^2}{a^2}, \quad \text{and} \tag{2}$$

$$\frac{\ddot{a}}{a} = -\frac{4\pi G}{3}\left(\rho + \frac{3p}{c^2}\right) + \frac{\Lambda c^2}{3}, \tag{3}$$

where $G$ is the gravitational constant, $\Lambda$ is the cosmological constant, $\rho$ is the mass density, $p$ is pressure, and $k$ is the spatial curvature. Following [8], Eq.(2) can be recast in terms of the effective kinetic energy and the effective potential:

$$\frac{1}{2}\dot{a}^2 + U(a) = E = -\frac{kc^2}{2}, \quad \text{where} \tag{4}$$

$$U(a) = -\frac{4\pi}{3}G\rho a^2 - \frac{\Lambda c^2 a^2}{6} \tag{5}$$

Assuming a dust-like equation of state, leading to $\rho a^3 = \mu = const$, we may obtain the effective universe potential $U(a)$, which is shown in Fig.1. The maximum of the effective potential corresponds to the Einstein static universe (ESU), which appears to be unstable to small perturbations, and can therefore be perturbed into an accelerating emergent cosmology, or into a contracting singular universe. The effective universe potential formalism appears to be quite useful, since for any functional form of the effective potential $U(a)$, there is an equation of state $p=p(\rho)$ that will produce it.

Note that the role of $a(t)$ in Eq.(4) is somewhat similar to the role of the center of mass coordinate, which describes a classic trajectory of motion of some complicated macroscopic object. In the semi-classical quantum mechanical picture, we should expect $a(t)$ not to be well-defined anymore. Instead, we should expect $a(t)$ to experience quantum fluctuations. Since the mass of the universe is very large, the position-momentum uncertainty principle written as $\Delta a \Delta p > h$ would lead to relatively small $\Delta a$. On the other hand, it is expected that quantum gravity should lead to modification of the conventional uncertainty principle, leading to quantum fluctuations of position variables at some minimum length scale, which is independent of object's mass. It is broadly accepted that this minimum length scale coincides with the Planck length $l_p=(hG/2\pi c^3)^{1/2}$ (see for example [9]), so that the expected quantum uncertainty of the scale factor is at least $\Delta a > l_p$. Somewhat similar ultra-fast quantum fluctuations of the scale factor may also be expected within the scope of the stochastic interpretation of quantum mechanics [10].

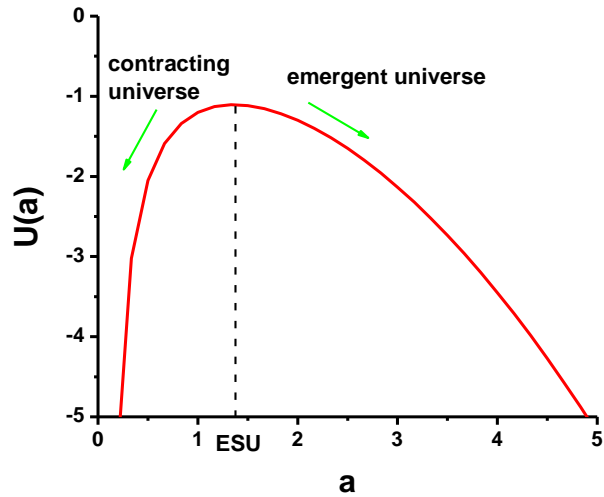

**Figure 1.** Effective potential of the universe $U(a)$ which is filled with dust-like matter.

In Section 3 we will consider the effect of these very fast fluctuations of the scale factor on the slow cosmological evolution of $a(t)$ given by the Friedmann equations (Eqs.(2,3)). It appears that the resulting modified dynamics of the universe may exhibit some similarities with the well-known dynamics of the inverted (Kapitza) pendulum [11] – see Fig.2. When a rigid pendulum of mass $m$ and length $l$ is subjected to rapid vertical vibrations at its pivot point, so that the motion of the pivot point may be described as $y_0 = A\cos(\omega t)$, the equation of motion for the angle $\theta$ may be obtained from the Euler-Lagrange equation as

$$\ddot{\theta} = \frac{\sin\theta}{l}\left(g - A\omega^2 \cos\omega t\right) \qquad (6)$$

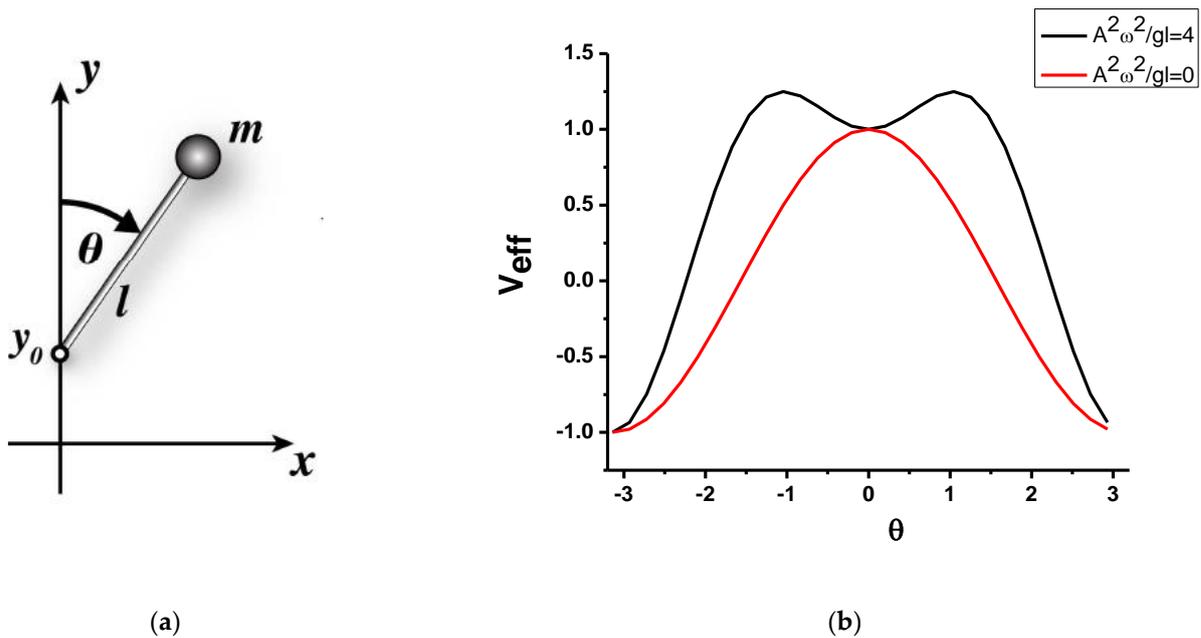

(a)  (b)

**Figure 2.** (**a**) A rigid pendulum with a vibrating pivot; (**b**) Effective potential $V_{eff}$ (in units of $mgl$) for an inverted (Kapitza) pendulum for the cases of $A^2\omega^2/gl = 0$ and $A^2\omega^2/gl = 4$.

Following the Kapitza's suggestion [11], the temporal evolution of the angular variable $\theta$ of such a system may be decomposed into the sum of fast and slow-varying variables:

$$\theta(t) = \theta_s(t) + \theta_f(t) , \qquad (7)$$

where $\theta_s$ is a slow varying function over one oscillation cycle and $\theta_f$ is the rapidly oscillating component of the angle $\theta$. Based on Eq.(6), the rapidly oscillating component is given by

$$\theta_f = \frac{A\sin\theta}{l}\cos\omega t \qquad (8)$$

After averaging over the rapid oscillations, the equation of motion for the slow component may be derived from an effective potential

$$V_{eff} = mgl\left(\cos\theta + \frac{A^2\omega^2}{4gl}\sin^2\theta\right) , \qquad (9)$$

which is obtained by adding the kinetic energy of rapid oscillations to the potential energy of slow motion. The resulting effective potential may exhibit two stable minima, as illustrated in Fig.2b. The originally unstable upward position $\theta=0$ (at $A=0$) becomes stable, due to the fact that the pendulum is subjected to rapid oscillations of the pivot point.

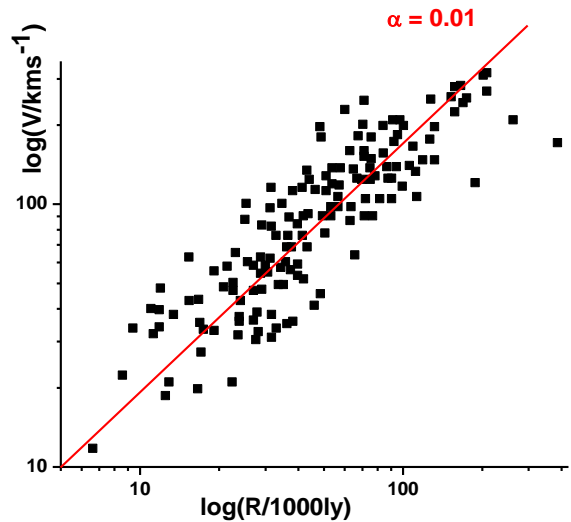

**Figure 3.** Log–log plot of the observed circular velocity $V$ as a function of the maximum radius $R$ for low-redshift disk galaxies re-plotted using data from Figure 4 of Ref. [12]. Low scatter around the "universal slope" $1.84 \times 10^{-3}$ km/s/ly indicated by the red line implies that this relationship is generic for disc galaxies in the low-redshift universe. As a result, the galaxies behave as "cosmic clocks", rotating roughly once a billion years at the very outskirts of their discs. This "universal slope" corresponds to $\alpha \sim 0.01$ [7].

Given the simple example above, we may expect that the rapid oscillations of the cosmological scale factor $a(t)$ may potentially alter the dynamics of galaxies and the universe itself. Indeed, it was demonstrated very re-cently that fast scale factor oscillations modify the current sign and magnitude of the average cosmological force, thus making it one of the potential probable causes for the modification of Newtonian dynamics in galaxy-scale systems [7]. In the presence of cosmological scale factor oscillations, the expected velocity distribution of test bodies far from the compact central mass *M* becomes

$$v = \sqrt{\frac{GM}{r} + \frac{\alpha c^2}{a^2} r^2} \quad , \qquad (10)$$

where $\alpha \sim 1$ is some dimensionless constant and *a* is the current magnitude of the "slow" scale factor [7]. As illustrated in Fig.3, this prediction replicates the recently discovered "cosmic clock" behavior of disk galaxies in the low-redshift universe [12], and it does not require the presence of dark matter. Given this recent result, it should also be interesting to consider the effect the fast fluctuations of the cosmological scale factor might play in the macroscopic evolution of the universe on the slow cosmological scales.

We should also note that the oscillations considered above do not need to be classical in nature. In fact, recent experimental and theoretical work established that purely quantum, random and incoherent fluctuations cause similar effects in dynamical systems which may be described by the Kapitza inverted pendulum model. Very recent result by Chawla et al. [13] demonstrates that the behavior of a quantum inverted pendulum under the influence of quantum fluctuations is similar to the behavior of the classical Kapitza pendulum where the point of support is vibrated vertically with a frequency higher than the critical value needed to stabilize its inverted position. Moreover, this phenomenon has been observed in an actual experiment in the non-equilibrium dynamics of a spin - 1 Bose-Einstein Condensate [14].

## 3. Results

Let us now follow the Kapitza prescription, and decompose the temporal evolution of the universe scale factor *a(t)* into the slow cosmological part and the fast-fluctuating part:

$$a(t) = a_s(t) + a_f(t) = a_s(t) + a_f \sin \omega t \quad , \qquad (11)$$

where for the sake of simplicity the fast-fluctuating part of *a(t)* is assumed to follow a simple sinusoidal time dependence (note that the quantum mechanical treatment of this effect looks virtually identical to the classical treatment – see ref. [13]). Based on the quantum mechanical uncertainty principle, the amplitude $a_f$ of these fast fluctuations must depend on $a_s$. By taking into account the generally assumed modifications of the uncertainty principle due to quantum gravity [9], we may expect the following simplified functional form of $a_f(a_s)$ dependence:

$$a_f = \frac{\eta}{a_s} + \delta \quad , \qquad (12)$$

where the first term corresponds to the conventional Heisenberg uncertainty principle (since $a_f \propto \dot{a}_f$ proportionality may be assumed for the fast scale factor fluctuations), and the second term corresponds to the quantum gravity corrections due to minimum length, so that $\delta \sim l_p$ [9]. Substituting Eq.(11) into Eqs.(2,3) (and neglecting the higher order terms) we obtain:

$$\frac{(\dot{a}_s + a_f \omega \cos \omega t)^2}{(a_s + a_f \sin \omega t)^2} = \frac{8\pi}{3} G\rho + \frac{\Lambda c^2}{3} - \frac{kc^2}{a^2}, \quad \text{and} \qquad (13)$$

$$\frac{\ddot{a}_s - a_f \omega^2 \sin \omega t}{a_s + a_f \sin \omega t} = -\frac{4\pi G}{3}\left(\rho + \frac{3p}{c^2}\right) + \frac{\Lambda c^2}{3} \qquad (14)$$

Using Eq.(12), we may transform Eqs.(13,14) as

$$\frac{\left(\dot{a}_s + \left(\frac{\eta}{a_s}+\delta\right)\omega\cos\omega t\right)^2\left(1+\left(\frac{\eta}{a_s^2}+\frac{\delta}{a_s}\right)\sin\omega t\right)^{-2}}{a_s^2} = \frac{8\pi}{3}G\rho + \frac{\Lambda c^2}{3} - \frac{kc^2}{a^2}, \text{ and} \qquad (15)$$

$$\frac{\left(\ddot{a}_s - \left(\frac{\eta}{a_s}+\delta\right)\omega^2\sin\omega t\right)\left(1+\frac{1}{a_s}\left(\frac{\eta}{a_s}+\delta\right)\sin\omega t\right)^{-1}}{a_s} = -\frac{4\pi G}{3}\left(\rho+\frac{3p}{c^2}\right) + \frac{\Lambda c^2}{3} \qquad (16)$$

Finally, by separating fast and slow variables in Eqs.(15,16), we obtain the following set of modified Friedmann equations:

$$\frac{\dot{a}_s^2}{a_s^2} = \frac{8\pi}{3}G\rho_s + \frac{\Lambda c^2}{3} - \frac{kc^2}{a_s^2} - \left(\frac{\eta}{a_s}+\delta\right)^2 \frac{\omega^2}{2a_s^2}, \text{ and} \qquad (17)$$

$$\frac{\ddot{a}_s}{a_s} = -\frac{4\pi G}{3}\left(\rho_s + \frac{3p_s}{c^2}\right) + \frac{\Lambda c^2}{3} - \left(\frac{\eta}{a_s}+\delta\right)^2 \frac{\omega^2}{2a_s^2}, \qquad (18)$$

where the slow parts of density $\rho_s$ and pressure $p_s$ will need to be determined based on the functional forms of $\rho(a)$ and $p(a)$ dependencies, and where we have kept only the lowest order terms in $1/a_s$ while eliminating all the fast-oscillating terms. However, if an analysis at early cosmological times is desirable, higher order terms in $1/a_s$ in these equations may be recovered from Eqs.(15,16) in a straightforward fashion.

As expected from the discussion in Section 2, fast oscillations of the cosmological scale factor do indeed considerably alter universe dynamics on the slow cosmological scales. For example, a quasi-static cosmological solution may be constructed even in the case of "empty" universe having $\rho_s=0$ and $p_s=0$. Such an unusual solution, having near zero first and second derivatives of $a_s(t)$, may be obtained if $k=0$ and $\Lambda=1.5a_s^{-2}$ (assuming that $a_s$ is large). Note that $a_f^2\omega^2/c^2\sim 1$ condition should be expected for quantum oscillations of the universe scale factor on the Planck scale.

Another interesting consequence of Eqs.(17,18) is that the presence of fast cosmological scale factor oscillations generically reproduces the late-time accelerated expansion of the universe regardless of its matter and energy content. This is quite obvious from Eq.(18), in which the late time acceleration is defined by the balance between the cosmological constant term and the $\delta^2\omega^2/2a_s^2$ term. Based on Eq.(18), the cosmological constant term starts to dominate at sufficiently large $a_s$. The observed "effective $\Lambda$" in the current epoch, which may be characterized as the beginning of an accelerated expansion phase, would be of the order of $\Lambda\sim 1/a^2$ if $a_f^2\omega^2/c^2\sim 1$ is assumed, which is consistent with the currently measured value of $\Lambda=1.11 \times 10^{-52}$ m$^{-2}$ [2]. It is noteworthy that the oscillating scale factor hypothesis is capable of reproducing both the experimentally observed modification of galaxy dynamics (see Fig.3), and the magnitude of the currently observed $\Lambda$ with the same choice $a_f^2\omega^2/c^2\sim 1$ of the oscillation parameters, which is consistent with the expected quantum fluctuations of $a(t)$ on the Planck scale.

Based on Eq.(17), we may also observe that similar to the case of Kapitza pendulum, the effective potential of the universe is strongly modified under the influence of quantum scale factor fluctuations. Indeed, the modified expression for the effective universe potential may be written as:

$$U_{eff}(a_s) = -\frac{4\pi}{3}G\rho_s a_s^2 - \frac{\Lambda c^2 a_s^2}{6} + \left(\frac{\eta}{a_s} + \delta\right)^2 \frac{\omega^2}{4} \qquad (19)$$

As we have discussed above, on the long cosmological time scales the effect of fluctuations leads to late-time acceleration of universe expansion. Let us now concentrate on what may happen with $U_{eff}(a)$ on the short time scales, and demonstrate that corrections due to quantum scale factor fluctuations may indeed considerably modify the initial dynamics of the universe.

As usual [15], let us assume that the density term in Eq.(2) consists of radiation density and matter density contributions, so that we may write it as

$$\rho = \frac{\mu}{a^3} + \frac{\xi}{a^4} \qquad (20)$$

Substituting Eq.(20) into Eq.(19) and neglecting higher order terms, we may obtain the following modified expression for the effective potential:

$$U_{eff}(a_s) = \left(-\frac{4\pi}{3}G\xi + \frac{\eta^2\omega^2}{4}\right)\frac{1}{a_s^2} + \left(-\frac{4\pi}{3}G\mu + \frac{\eta\delta\omega^2}{2}\right)\frac{1}{a_s} + \frac{\delta^2\omega^2}{4} - \frac{\Lambda c^2 a_s^2}{6} =$$

$$= C_{-2}\frac{1}{a_s^2} + C_{-1}\frac{1}{a_s} + C_0 - \frac{\Lambda c^2 a_s^2}{6} \qquad (21)$$

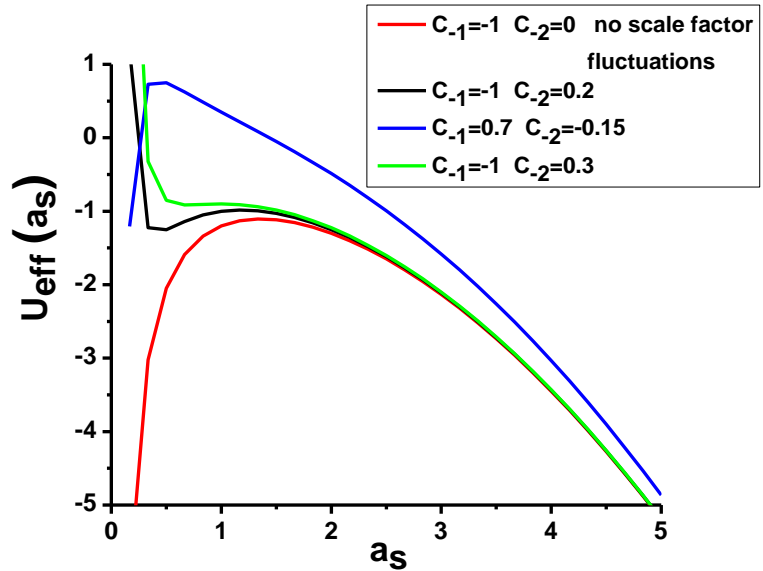

**Figure 4.** Variations of the effective potential of the universe $U_{eff}(a_s)$ under the influence of scale factor fluctuations assuming different matter and radiation content. The original potential calculated without scale factor fluctuations (see Fig.1) is shown by the red line.

Depending on the matter and radiation content of the universe, the shape of $U_{eff}$ may now exhibit considerable variations compared to the potential shown in Fig.1, which was derived in the absence of quantum scale factor fluctuations. Some examples of these variations are depicted in Fig.4. Similar to the inverted Kapitza pendulum, a stable local potential energy minimum may appear near the initially unstable static universe solution under the influence of quantum scale factor fluctuations – see for example black and green curves in Fig.4. In the scenario represented by the black curve, the universe may originate due to quantum tunneling from a metastable state at $a_s$~0.4 into an accelerating state at $a_s$>2, thus giving rise to an emergent inflationary cosmology.

## 4. Discussion and Conclusions

As we have seen from the detailed analysis in Section 3, quantum fluctuations of the scale factor should modify the Friedmann equations, leading to considerable modification of cosmological evolution. The late-time accelerated expansion of the universe appears to be a generic feature of such modified cosmological evolution. It also appears that quantum fluctuations of the universe scale factor may considerably modify the universe potential (see Fig.4), which indicates importance of these corrections for the ultimate theory of inflation. Indeed, the examples considered above provide a toy model for the emergent universe scenario in which inflation takes place eternally. A more realistic scenario can be constructed if one replaces $\Lambda$ by the inflaton field. The inflaton field, originally postulated by Alan Guth,[16] provides a mechanism by which a period of rapid inflationary expansion can be generated, forming a universe consistent with the observed spatial isotropy and homogeneity. A spatially homogeneous scalar inflaton field minimally coupled to gravity is described by the Lagrangian density

$$L = \frac{1}{2}\dot{\phi}^2 - V(\phi) \qquad (22)$$

The energy density and pressure of such a field are given, respectively, by

$$\rho_\phi = \frac{1}{2}\dot{\phi}^2 + V(\phi) , \qquad (23)$$

and

$$p_\phi = \frac{1}{2}\dot{\phi}^2 - V(\phi) , \qquad (24)$$

which may be substituted into the Friedmann equations (Eqs.(2,3)) to obtain the cosmological evolution of $a(t)$ and the effective universe potential $U(a)$. Based on Eq.(3), it is clear that the strong energy condition (SEC) $\rho+3p \geq 0$ must be violated in order to realize inflation scenarios. The cosmological constant with $p = -\rho$ presents us with an example of such violation. When the $\Lambda$ term in the simplified models considered in Section 3 is replaced with any kind of combination of the inflaton field and $\Lambda$, we will still need to take into account the influence of scale factor fluctuations. Based on our results, it is clear that the functional form of the desired inflaton potential may be simplified, since such necessary features of inflation theory as slow roll and reheating may potentially result from the scale factor fluctuations – see for example the green line in Fig.4.

We should also note that the effective universe potential in the quantum Friedmann models exhibits a similar unstable point (see for example ref.[17]). Therefore, it is reasonable to expect that random and incoherent quantum fluctuations of the scale factor should lead to Kapitza effect, especially since there is no reason to expect that the quantum state of the universe within the scope of the quantum Friedmann model will be a scale factor eigenstate.

In addition, as we mentioned above, the quantum inverted pendulum dynamics has been observed in an actual experiment in the non-equilibrium dynamics of a spin - 1 Bose-Einstein Condensate [14]. When information theoretic methods are applied to quantum spin chain dynamics, a Riemannian metric may typically be defined on the equilibrium thermodynamic state space of such a system [18]. In the Lipkin-Meshkov-Glick model, such systems exhibit a quantum phase transition, which is very similar to the inverted pendulum dynamics, so it will be interesting to use such an experimentally observed spin - 1 Bose-Einstein Condensate system as a "quantum simulator" of the universe.

**References**

1. Wang, Q.; Zhu, Z.; Unruh, W.G. How the huge energy of quantum vacuum gravitates to drive the slow accelerating expansion of the Universe. *Phys. Rev. D* **2017**, *95*, 103504.
2. Planck Collaboration Planck 2013 results. XXXII. The updated Planck catalogue of Sunyaev-Zeldovich sources. *Astron. Astrophys.* **2014**, 571, A16.
3. Matsui, H.; Watamura, N. Quantum spacetime instability and breakdown of semiclassical gravity. *arXiv* **2020**, arXiv:1910.02186.
4. Pace, F.; Fedeli, C.; Moscardini, L.; Bartelmann, M. Structure formation in cosmologies with oscillating dark energy. *Mon. Not. R. Astron. Soc.* **2012**, *422*, 1186–1202.
5. Zaripov, F. Oscillating cosmological solutions in the modified theory of induced gravity. *Adv. Astron.* **2019**, *2019*, 1502453.
6. Ringermacher, H. I.; Mead, L.R. Observation of discrete oscillations in a model-independent plot of cosmological scale factor versus lookback time and scalar field model. *Astron. J.* **2015**, *149*, 137.
7. Smolyaninov, I. I. Oscillating cosmological force modifies Newtonian dynamics. *Galaxies* **2020**, *8*, 45.
8. Sahni, V.; Starobinsky, A. A. The case for a positive cosmological lambda term, *Int. J. Mod. Phys. D* **2000**, *9*, 373.
9. Garay, L. J. Quantum gravity and minimum length. *Int. J. Mod. Phys. A* **1995**, *10*, 145.
10. Davidson, M. P. The origin of the algebra of quantum operators in the stochastic formulation of quantum mechanics. *Letters in Mathematical Physics* **1979**, *3*, 367.
11. Kapitza, P.L. Dynamic stability of a pendulum when its point of suspension vibrates. *Soviet Phys. JETP* **1951**, *21*, 588.
12. Meurer, G.R.; Obreschkow, D.; Wong, O.I.; Zheng, Z.; Audcent-Ross, F.M.; Hanish, D.J. Cosmic clocks: A tight radius-velocity relationship for HI-selected galaxies. *Mon. Not. R. Astron. Soc.* **2018**, *476*, 1624–1636.
13. Chawla, R.; Paul, S.; Bhattacharjee, J.K. Quantum fluctuations stabilize an inverted pendulum. *arXiv* **2019,** arXiv:1904.00975 [nlin.CD].
14. Gerving, C.S.; Hoang, T.M.; Land, B.J.; Anquez, M.; Hamley, C.D.; Chapman, M.S. Non-equilibrium dynamics of an unstable quantum pendulum explored in a spin-1 Bose Einstein condensate. *Nature Communications* **2012**, *3*, 1169.
15. Bag, S.; Sahni, V.; Shtanov, Y.; Unnikrishnan, S. Emergent cosmology revisited. *J. Cosmology and Astroparticle Phys.* **2014**, 07, 034.
16. Guth, A. H. Inflationary universe: A possible solution to the horizon and flatness problems. *Phys. Rev. D* **1981**, 23, 347.
17. Siong, C.H.; Radiman, S.; Nikouravan, B. Friedmann equation with quantum potential. AIP Conference Proceedings **2013**, *1571*, 35.
18. Dey, A.; Mahapatra, S.; Roy, P.; Sarkar, T. Information geometry and quantum phase transitions in the Dicke model. *Phys. Rev. E* **2012**, 86, 031137.